\documentstyle[12pt,fleqn]{article}
\newcommand{\be}{\begin{equation}}
\newcommand{\ee}{\end{equation}}
\newcommand{\bea}{\begin{eqnarray}}
\newcommand{\eea}{\end{eqnarray}}

\catcode`@=11
\@addtoreset{equation}{section}
\catcode`@=12
\mathchardef\SGamma="7100
\begin{document}
\begin{titlepage}
\begin{flushright}
Freiburg THEP-98/27\\
gr-qc/9901055
\end{flushright}
\vskip 1cm
\begin{center}

{\large\bf  Effective action and decoherence by fermions in
            quantum cosmology}
\vskip 1cm
{\bf Andrei O. Barvinsky}
\vskip 0.4cm
Theory Department, \\ Levedev Institute and Lebedev Research Center in Physics,\\
Leninsky Prospect 53, Moscow 117924, Russia.

\vskip 0.7cm
{\bf Alexander Yu. Kamenshchik}
\vskip 0.4cm
L.D. Landau Institute for Theoretical Physics,\\ Russian Academy of Sciences,\\
Kosygin Street 2, Moscow 117334, Russia.

\vskip 0.7cm
{\bf Claus Kiefer}
\vskip 0.4cm
 Fakult\"at f\"ur Physik, Universit\"at Freiburg,\\
  Hermann-Herder-Stra\ss e 3, D-79104 Freiburg, Germany.

\end{center}
\vskip1cm
\small
\begin{center}
{\bf Abstract}
\end{center}
\begin{quote}
We develop the formalism for the one-loop no-boundary state in a cosmological model
with fermions. We use it to calculate the reduced density matrix for
an inflaton field by tracing out the fermionic degrees of freedom, yielding
both the fermionic effective action and the standard decoherence factor.
We show that dimensional regularisation of ultraviolet divergences
would lead to an inconsistent density matrix.
Suppression of these divergences to zero is instead performed
through a nonlocal Bogoliubov transformation of the fermionic variables,
which leads to a consistent density matrix.
The resulting degree of decoherence is less than in the case of bosonic fields.
\end{quote}

\end{titlepage}

\section{Introduction}
\hspace{\parindent}
Quantum cosmology -- the application of quantum theory to the Universe
as a whole -- is  undergoing now a period of successful development connected
with the fact that it can help to explain the origin of inflation
in the early Universe.
One of the basic problems in quantum cosmology is the choice of an
appropriate boundary condition. Two proposals have become especially
popular: The
no-boundary state \cite{Hartle-Hawk,Hawk} and the tunneling state \cite{tun}.
Analysis of these two proposals in the one-loop approximation allows one
to formulate the conditions for the normalisability of the wave function of
the Universe \cite{norm} and to calculate the probability peak for the
energy scale of inflation in different models \cite{tunnel,qsi}. Such a
probability peak has been interpreted as providing a criterium to select
amongst the members of an ``ensemble" of classical universes.

However, quantum theory does not yield a classical ensemble.
A complete analysis would thus have to include a quantitative discussion
of the quantum-to-classical transition.

It is now generally accepted that the classical properties for a subsystem
arise from the irreversible interaction of this quantum system with its
natural environment. Starting with the pioneering work of Zeh
in the seventies \cite{Zeh}, this concept of {\em decoherence}
has been developed extensively, see \cite{decoherence1} and
\cite{decoherence2} for reviews. Quite recently, this continuous
loss of coherence was observed in quantum-optical experiments
\cite{Haroche}.

Quantum-to-classical transition through decoherence also applies to quantum
cosmology where it was suggested to consider background degrees of freedom
(such as the scale factor and the homogeneous mode of the inflaton scalar
field) as the relevant system and the perturbations (such a gravitational
waves or density fluctuations) as an environment \cite{Zeh2}. A
quantitative discussion of this idea was first done in \cite{Kiefer1},
where it was demonstrated how the scale factor and the inflaton field can
acquire classical properties. This idea was further pursued in many papers
\cite{decoherence1}. The quantum-to-classical transition through
decoherence plays also a crucial role for the primordial fluctuations
itself that eventually serve as classical seeds for galaxies and clusters
of galaxies \cite{Polarski}.

However, there is one big problem arising in the application of this idea
to cosmology. Since there are infinitely many environmental modes, there
arise ultraviolet divergences that have to be removed. It was already noted
in \cite{Paz} that dimensional regularisation is only applicable for the
phase part of decoherence factor, not its absolute value. In our preceding
paper \cite{we-decoh} this was confirmed by detailed calculations. It was also shown in
\cite{we-decoh}, that for the case of bosonic fields, using the so-called
conformal parametrisation one can get rid of ultraviolet divergences and
obtain finite results for the absolute value of the decoherence factor. An
analogous idea was investigated in QED and in a particular cosmological
model in \cite{Claus-elec}. A similar proposal was also discussed in
\cite{Okamura}.

The purpose of the present paper is to extend the analysis of
\cite{we-decoh} to fermionic degrees of freedom. From a physical point of
view, this is important because fermions constitute the main part of physical
fields in our Universe. On the other hand, the Grassmannian character of
fermion variables implies an essential modification of the formalism which
is of interest by itself.

First of all, we want to emphasise that the formalism of the Hartle-Hawking wave
function of the Universe via a path integral {\em in one-loop
approximation} was previously developed only for bosonic fields
\cite{norm,tunnel,qsi}. Here, we extend this formalism to fermionic
fields. In this case the structure of the norm of partial wave functions,
describing perturbations of different harmonics of fermionic field - an
important ingredient of the one-loop formalism - is more complicated than
in the case of bosonic fields where it can be reduced to
calculating the Wronskian of the corresponding basis functions.
Using this formalism, we can then calculate the decoherence factor for
the reduced density matrix of scale factor and inflaton field.
However, like in the bosonic case, ultraviolet divergences appear,
as was already noted in an early paper on decoherence due
to fermions \cite{Claus-ferm}.

By handling the divergent series within dimensional regularisation
\cite{dimen,Bar-Vil}, we obtain a renormalised expression for the
absolute value of the decoherent factor. Unfortunately, however, it shows a wrong
behaviour violating the consistency conditions for a density matrix as a
bounded operator, let alone its decoherence properties. A similar
phenomenon was observed also in the case of bosonic fields \cite{we-decoh}.
However, a simple reparametrisation of field variables by overall
multiplication with an appropriate power of the scale factor allowed us
there to obtain finite expressions. This does not work in the case of
fermionic fields, where such a reparametrisation is incompatible with the
structure of the measure on the corresponding Hilbert space of states.
However, there exists another possibility: Performing a Bogoliubov
transformation of field variables one can get again finite expressions for
the reduced density matrix. It is shown that such Bogoliubov
transformations are in a certain sense analogous to the well-known
Foldy-Wouthuysen transformation of spinor variables \cite{Itzyk-Zub}. Thus,
as in the case of bosonic fields, the definition of the environment is
crucial for the consistency and decoherence properties of the reduced
density matrix, but in the fermionic case this definition is much more
involved.

Our paper is organised as follows. In section 2 we briefly review the
spinor formalism for quantum cosmology. This section includes the relation
between the two- and four-component spinor formalism and presents the
expansion of Weyl spinors in three-dimensional spinor harmonics with a
special emphasis on the dimensionalities of the $O(d)$ spinor
representation on $d$-dimensional spheres (necessary for dimensional
regularisation). Section 3 is devoted to the quantisation of fermion fields
in the Hamiltonian formalism and to the construction of the cosmological
wave function. In section 4 we discuss the connection between path
integral, Euclidean effective action and density matrix of fermions.
Section 5 contains the presentation of contributions of massless and
massive fermions into the amplitude of the decoherence factor in the
reduced density matrix. In section 6 we demonstrate that the application of
dimensional regularisation violates basic properties of the density matrix.
Section 7 comprises the discussion of Bogoliubov transformation of
fermionic variables, allowing to get rid of ultraviolet divergences in the
decoherence factor. We also exhibit the connection with the traditional
Foldy-Wouthuysen transformations. Section 8 contains a brief summary of the
obtained results.

\section{Spinor formalism for quantum cosmology}
\hspace{\parindent}
In the main part of this paper we shall use a two-component spinor
formalism as it is usually done in  papers with applications to quantum
cosmology \cite{Hall-ferm,we-Kluw,Esp-Camb}. However, at the beginning we
would like to present some formulae connecting this formalism with that of
four-dimensional Dirac spinors (see, for example Ref. \cite{Wess}).

We shall use two-component Weyl spinors in Van der Waerden notation. There
are two types of spinors with primed and unprimed indices which are
transformed under the action of matrix $M \in SL(2,C)$ according to the
following rules:
          \begin{eqnarray}
          &&\psi'_{A} = M_{A}^{B} \psi_{B},\nonumber \\
          &&\psi'^{A} = (M^{-1})^A_{B} \psi^{B},\nonumber \\
          &&\bar{\psi}'_{\dot{A}} =
          M^{* \dot{B}}_{\dot{A}}\bar{\psi}_{\dot{B}}\nonumber \\
          &&\bar{\psi}'^{\dot{A}} =
          (M^{-1})^{*\dot{A}}_{\dot{B}}\bar{\psi}^{\dot{B}}.
          \label{van-der}
          \end{eqnarray}
Those with dotted indices transform under the $\left(0,\frac12\right)$
representation of the Lorentz group, while those with undotted indices
transform under the $\left(\frac12,0\right)$ conjugate representation. For
the Minkowski metric we use the signature
          \begin{equation}
          \eta_{mn} = {\rm diag} (-1,1,1,1).
          \label{signature}
          \end{equation}
The Pauli matrices $\sigma_{A \dot{A}}^{m}$ have the following form:
          \begin{equation}
          \sigma^0 = \left(
          \begin{array}{rr}
          -1\  &\  0   \\
          0\  &\  -1
          \end{array}
          \right),
          \sigma^1 = \left(
          \begin{array}{rr}
          0\  &\  1   \\
          1\  & \ 0
          \end{array}
          \right),
          \sigma^2 = \left(
          \begin{array}{rr}
          0\  &\  -i   \\
          i\  &\  0
          \end{array}
          \right),
          \sigma^3 = \left(
          \begin{array}{rr}
          1\  &\  0   \\
          0\  &\  -1
          \end{array}
          \right).
          \end{equation}
Indices can be lowered and raised by tensors $\varepsilon_{AB}$ and
$\varepsilon_{\dot{A}\dot{B}}$ with components
           \begin{eqnarray}
           &&\varepsilon_{11} = \varepsilon_{22} =
            \varepsilon^{11} = \varepsilon^{22} = 0, \nonumber \\
           &&\varepsilon_{21} = \varepsilon^{12} = 1, \
           \varepsilon_{12} = \varepsilon^{21} = -1.
           \label{epsilon}
           \end{eqnarray}
(The components of  $\varepsilon$ with dotted indices are the same). It is
convenient also to define
           \begin{equation}
           \bar{\sigma}^{m \dot{A} A} =
           \varepsilon^{\dot{A}\dot{B}} \varepsilon^{AB}
           \sigma^{m}_{B \dot{B}}.
           \label{sigma-bar}
           \end{equation}
The anticommutation relations for the Pauli matrices are:
           \begin{eqnarray}
           &&(\sigma^m \bar{\sigma}^n + \sigma^n\bar{\sigma}^m)_A^B
           = -2\eta^{mn}\delta_A^B,\nonumber \\
           &&(\bar{\sigma}^m \sigma^n + \bar{\sigma}^n \sigma^m)
           _{\dot{B}}^{\dot{A}} = -2\eta^{mn}\delta_{\dot{B}}
           ^{\dot{A}}.
           \label{anticom}
           \end{eqnarray}

For four-dimensional $\gamma$ - matrices and Dirac spinors we shall use the
Weyl basis. In this basis, $\gamma$ - matrices are given
           \begin{equation}
           \gamma^m = \left(
           \begin{array}{ll}
           0\  &\  \sigma^m \\
           \bar{\sigma}^m\  &\  0
           \end{array}
           \right),
           \label{gamma-define}
           \end{equation}
while the Dirac spinor is related to the two-component Weyl spinors by
           \begin{equation}
           \Psi = \left(
           \begin{array}{c}
           \chi_A \\
           \bar{\phi}^{\dot{A}}
           \end{array}
           \right).
           \label{Dir-define}
           \end{equation}

It is worth noting that the canonical basis for the Dirac $\gamma$ -
matrices,
           \begin{equation}
           \gamma_C^0 = \left(
           \begin{array}{rr}
           -1\  &\  0 \\
           0\  &\  1
           \end{array}\right), \
           \gamma_C^k = \left(
           \begin{array}{cc}
           0\  &\  \sigma^k \\
           -\sigma^k\  &\  0
           \end{array}\right),
           \label{can-bas}
           \end{equation}
is related to the Weyl basis (\ref{gamma-define}) by the following
similarity transformation:
            \begin{equation}
            \Gamma_W = X \Gamma_C X^{-1},\
            X = \frac{1}{\sqrt{2}}
            \left(\begin{array}{rr}
            1 & -1 \\
            1 & 1
            \end{array}\right).
            \label{similarity}
            \end{equation}
Note that the Weyl basis is more consistent from the group-theoretical
point of view because in this basis the Dirac spinor is represented as a
direct sum of two spinors belonging to irreducible representations of the
Lorentz group (or to be precise, its covering group $SL(2,C)$). This basis
is also more convenient for the description of the massless limit for
ultrarelativistic fermions. At the same time, the canonical basis is
convenient for studying the non-relativistic limit and for the transition
from the Dirac equation to the Pauli equation. We shall come back to this
question when considering the relation between Bogoliubov transformations
and the Foldy-Wouthuysen representation for Dirac spinors.

Now, starting with the traditional action for Dirac spinors in
four-dimensional spacetime \cite{Bog-Shir, Itzyk-Zub} and using the above
relations (\ref{gamma-define}),(\ref{Dir-define}), one can get the
following action \cite{Hall-ferm}:
            \begin{equation}
            S = -\frac{i}{2}\int d^4x e (\bar{\phi}^{\dot{A}}
            e^{\mu}_{A\dot{A}}D_{\mu}\phi^A
            + \bar{\chi}^{\dot{A}}
            e^{\mu}_{A\dot{A}}D_{\mu}\chi^A)
            + h.c. - m\int d^4x e (\chi_A \phi^A
            + \bar{\phi}^{\dot{A}}\bar{\chi}_{\dot{A}}),
            \label{action}
            \end{equation}
where $D_{\mu}$ is a spinorial covariant derivative,
$e^{\mu}_{A\dot{A}}$ is the Pauli matrix contracted with the tetrad
$e^{\mu}_m$, and $e$ is the determinant of the tetrad.

Varying the action (\ref{action}), one can get the Dirac equations
            \begin{eqnarray}
            &&e^{\mu}_{A\dot{A}}D_{\mu}\phi^A = i m\bar{\chi}_{\dot{A}},
            \nonumber \\
            &&e^{\mu}_{A\dot{A}}D_{\mu}\chi^A = i m\bar{\phi}_{\dot{A}},
            \nonumber \\
            &&e^{\mu}_{A\dot{A}}D_{\mu}\bar{\phi}^{\dot{A}} =
            -i m\bar{\chi}_A,
            \nonumber \\
            &&e^{\mu}_{A\dot{A}}D_{\mu}\bar{\chi}^{\dot{A}} =
            -i m\bar{\phi}_A.
            \label{Dirac-eq}
            \end{eqnarray}

In what follows we shall consider fermionic perturbations on the background
of a spatially closed Friedmann model with $S^3$ spatial sections. We shall
work within the inflationary model with the spacetime metric \cite{Hawk}
            \begin{equation}
            ds^2 = -dt^2 + a^2(t) d\Omega_3^2,
            \label{metric}
            \end{equation}
($d\Omega_3^2$ is the metric on the unit three-sphere) with a cosmological
scale factor evolving according to the law
            \begin{equation}
            a(t) = \frac{1}{H}\cosh Ht,
            \label{cosm-factor}
            \end{equation}
where $H = H(\varphi)$ is an effective Hubble constant possibly generated
by an inflaton scalar field $\varphi$.

Then one should expand the fermionic field using the full set of spinor
harmonics on the three-sphere \cite{Hall-ferm}:
            \begin{eqnarray}
            &&\phi_A = \frac{a^{-3/2}}{2\pi}
            \sum_{np}\sum_q\alpha_n^{pq}
            [m_{np}(t)\rho_A^{nq}(x) + \bar{r}_{np}(t)
            \bar{\sigma}_A^{nq}(x)],\nonumber \\
            &&\bar{\phi}_{\dot{A}} = \frac{a^{3/2}}{2\pi}
            \sum_{np}\sum_q\alpha_n^{pq}
            [\bar{m}_{np}(t)\bar{\rho}_{\dot{A}}^{nq}(x)
            + r_{np}(t)\sigma_{\dot{A}}^{nq}(x)],\nonumber \\
            &&\chi_A = \frac{a^{-3/2}}{2\pi}
            \sum_{np}\sum_q\beta_n^{pq}
            [s_{np}(t)\rho_A^{nq}(x) + \bar{t}_{np}(t)
            \bar{\sigma}_A^{nq}(x)],\nonumber \\
            &&\bar{\chi}_{\dot{A}} = \frac{a^{3/2}}{2\pi}
            \sum_{np}\sum_q\beta_n^{pq}
            [\bar{s}_{np}(t)\bar{\rho}_{\dot{A}}^{nq}(x) + t_{np}(t)
            \sigma_{\dot{A}}^{nq}(x)].
            \label{harmonics}
            \end{eqnarray}
Here, the weight factor $a^ {-3/2}$ is especially chosen in order to write
the action (\ref{action}) in a special parametrisation reflecting the
conformal properties of the spinor field. It is conformally invariant in
the massless case and, therefore, decouples in this parametrisation from
the gravitational background. Summation over $n$ runs from $1$ to $\infty$,
while the indices $p$ and $q$ run from $1$ to $n(n+1)$ - the dimensionality
of the irreducible representation for Weyl spinors. The time-dependent
coefficients $m_{np}, r_{np}, t_{np}, s_{np}$ and their complex conjugates
are taken to be odd elements of a Grassmann algebra. The constant
coefficients are included for convenience in order to avoid couplings between
different values of $p$ in the expansion of the action. They may be
regarded as block-diagonal matrices of dimension $n(n+1)$ whose structure
can be found in \cite{Hall-ferm}. The harmonics
$\rho,\sigma,\bar{\rho},\bar{\sigma}$ are normalised eigenfunctions of the
Dirac operator on the three-sphere with eigenvalues $\pm(n+1/2),n\geq 1$.
Here we are using a notation that differs from \cite{Hall-ferm} by a unit
shift of the quantum number $n$. Inserting (\ref{harmonics}) into the
action (\ref{action}) and using the properties of spinor harmonics, one
finds that the action reads
            \begin{eqnarray}
            &&S = \sum_{np}\int dt\left(\frac{i}{2}
            (\bar{m}_{np}\dot{m}_{np} + m_{np}\dot{\bar{m}}_{np}
            + \bar{s}_{np}\dot{s}_{np} + s_{np}\dot{\bar{s}}_{np}
            \right.\nonumber \\
            &&\qquad\qquad+ \bar{t}_{np}\dot{t}_{np} + t_{np}\dot{\bar{t}}_{np}
            + \bar{r}_{np}\dot{r}_{np} + r_{np}\dot{\bar{r}}_{np})
            \nonumber \\
            &&\qquad\qquad+ \frac1a (n+\frac12)(\bar{m}_{np}m_{np}
            + \bar{s}_{np}s_{np} + \bar{t}_{np}t_{np}
            + \bar{r}_{np}r_{np}) \nonumber \\
            &&\qquad\qquad\left.- m(r_{np}t_{np} + \bar{t}_{np}\bar{r}_{np}
            + s_{np}m_{np} + \bar{m}_{np}\bar{s}_{np})
            {\phantom{\frac12}}\right)\nonumber\\
            &&\qquad\qquad+\frac i2\sum_{np}
            \left(\bar{m}_{np}{m}_{np} + \bar{s}_{np}{s}_{np}
            + \bar{t}_{np}{t}_{np} +
            \bar{r}_{np}{r}_{np}\right)_{S_{I}}\nonumber\\
            &&\qquad\qquad+\frac i2\sum_{np}
            \left(\bar{m}_{np}{m}_{np} + \bar{s}_{np}{s}_{np}
            + \bar{t}_{np}{t}_{np} +
            \bar{r}_{np}{r}_{np}\right)_{S_{F}}.                \label{action1}
            \end{eqnarray}
The last two lines in this equation represent the boundary terms at the
initial $(S_{I})$ and final $(S_F)$ hypersurfaces of the spacetime domain.
They have to be necessarily included in order to guarantee the consistency
of the variational problem for this action with {\em fixed unbarred
variables} at $S_F$ and {\em fixed barred variables} at $S_{I}$ -- the
choice of boundary conditions characteristic for the definition of the
fermionic wave function in the holomorphic representation \cite{Hall-ferm}.
Thus the full action has the form of a sum of fermionic actions
            \begin{eqnarray}
            &&S_n = \int dt \left(\frac{i}{2}(\bar{x}\dot{x}
            +x\dot{\bar{x}} + \bar{y}\dot{y}
            +y\dot{\bar{y}})+e^{-\alpha}(n+\frac12)(\bar{x}x
            + \bar{y}y) - m(yx + \bar{x}\bar{y})\right)\nonumber\\
            &&\qquad\qquad+\frac i2(\bar{x}x+\bar{y}y)_{S_F}
            +\frac i2(\bar{x}x+\bar{y}y)_{S_I}.                   \label{action2}
            \end{eqnarray}
with the Grassmanian variables $x$ and $y$ denoting $m_{np}$ and $s_{np}$
or $t_{np}$ and $r_{np}$, respectively.

{}From the action (\ref{action2}) the following field equations may be
derived:
            \begin{eqnarray}
            &&i\dot{x} + \nu x - m\bar{y} = 0,\\
            \label{Dirac-eq1}
            &&i\dot{\bar{x}} - \nu \bar{x} + my = 0,\\
            \label{Dirac-eq2}
            &&i\dot{y} + \nu y + m\bar{x} = 0,\\
            \label{Dirac-eq3}
            &&i\dot{\bar{y}} - \nu \bar{y} - mx = 0,
            \label{Dirac-eq4}
            \end{eqnarray}
where
             \begin{equation}
             \nu = \frac{n + 1/2}{a}.
             \label{nu-define}
             \end{equation}

In the rest of this section we give the formula for the dimensionality of
$O(4)$ irreducible representation for spinors on spheres of arbitrary even
dimensionality, which we shall need in the dimensional-regularisation
calculations below. For this we use the technique which for scalar, vector
and tensor harmonics on $S^3$ has been constructed by Lifshitz and
Khalatnikov \cite{Lif-Khal} and for spinor harmonics on $S^3$ is described
in \cite{Hall-ferm}.

Let us consider flat Euclidean four-space with the metric
               \begin{equation}
               ds^2 = \delta_{\mu\nu}dx^{\mu}dx^{\nu}.
               \label{Cartesian}
               \end{equation}
The metric may also be written in spherical coordinates $r,\chi,\theta,
\phi$:
                \begin{equation}
                ds^2 = dr^2 + r^2 d\Omega_3^2,
                \label{spherical}
                \end{equation}
where
                \begin{equation}
                d\Omega_3^2 = d\chi^2 + \sin^2\chi(d\theta^2
                + \sin^2\theta d\phi^2).
                \label{spherical1}
                \end{equation}

Harmonics on $S^3$ are then obtained by considering the following
homogeneous polynomials in Cartesian coordinates,
                \begin{equation}
                r^n\rho_A(\chi,\theta,\phi) =
                T_{AA_1\cdots A_{n-1}\dot{A}_1\cdots \dot{A}_{n-1}}
                x^{A_1\dot{A}_1}\cdots x^{A_{n-1}\dot{A}_{n-1}},
                \ \ n=1,\ldots,
                \label{spinor}
                \end{equation}
where
                \begin{equation}
                x^{A\dot{A}} = e_{\mu}^{A\dot{A}}x^{\mu}
                \label{spinor1}
                \end{equation}
and $T_{AA_1\cdots A_{n-1}\dot{A}_1\cdots \dot{A}_{n-1}}$ is a constant
spinor of rank $(2n - 1)$, totally symmetric in all its indices and hence
traceless
(in the metric $\varepsilon_{AB}$). It is obvious that the number of independent harmonics is
defined by the number of different choices of dotted and and undotted
indices which can acquire only two values: $0$ or $1$. For $n-1$ dotted
indices there are $n$ opportunities while for $n$ undotted indices there
are $(n+1)$ different choices. Thus, the dimensionality of irreducible
representation for Weyl spinors harmonics on $S^3$ is
                 \begin{equation}
                 {\rm dim}(n,4) = n(n+1).
                 \label{dimen}
                 \end{equation}

In the case of a spacetime with even dimensionality $d$, the dimensionality
of the corresponding Clifford algebra of Dirac $\gamma$ - matrices is equal
to $2^d$; correspondingly the number of components of Dirac spinors is
$2^{d/2}$, while the number of components of Weyl spinors equals
$2^{(d-2)/2}$. Thus, indices $A$ and $\dot{A}$ in the expression for
constant tensor $T$ take $2^{(d-2)/2}$ different values. In such a way the
problem of calculation of dimensionality of irreducible representation for
spinor harmonics reduces to the combinatorial counting of the number of
possible decomposition of the set of $n$ elements into the union  of
$2^{(d-2)/2}$ subsets. One has
                  \begin{equation}
                  {\rm dim}(n,d) = \frac{\Gamma\left(n + 2^{(d-2)/2}
                  \right)}{\Gamma\left(2^{(d-2)/2}\right)
                  \Gamma(n+1)}
                  \times
                  \frac{\Gamma\left(n + 2^{(d-2)/2} - 1
                  \right)}{\Gamma\left(2^{(d-2)/2}\right)
                  \Gamma(n)}.
                  \label{dimen1}
                  \end{equation}

\section{Quantisation of fermion fields in Hamiltonian formalism and
cosmological wave function}
\hspace{\parindent}
Following the Dirac procedure for systems with constraints, one may obtain
from the action (\ref{action2}) the Hamiltonian
                   \begin{equation}
                   H_n(x,\bar{x},y,\bar{y}) = \nu(x\bar{x}+y\bar{y})
                   + m(yx + \bar{x}\bar{y}),
                   \label{Hamiltonian}
                   \end{equation}
where $x,y,\bar{x},\bar{y}$ obey the Dirac-brackets relations
                   \begin{equation}
                   \{x,\bar{x}\}^* = -i,\ \
                   \{y,\bar{y}\}^* = -i.
                   \label{brackets}
                   \end{equation}
The total Hamiltonian is the sum of the background Hamiltonian, depending
on the inflaton scalar field and the fermionic Hamiltonian, representing
the sum of Hamiltonians (\ref{Hamiltonian}). The total Hamiltonian
vanishes, giving in such a way the Hamiltonian constraint:
                   \begin{equation}
                   H_0 + H_f = 0.
                   \label{constraint}
                   \end{equation}
In the quantum theory, this Hamiltonian constraint becomes the
Wheeler-DeWitt equation
                    \begin{equation}
                    [H_0 + H_f] |\Psi\rangle = 0,
                    \label{Wheeler-DeWitt}
                    \end{equation}
where $|\Psi\rangle$ is the wave function of the Universe, and the fermionic
Hamiltonian is obtained from its classical counterpart by replacing the
variables with operators satisfying the anticommutation relations
                     \begin{equation}
                     [x,\bar{x}]_+ = 1,\ \
                     [y,\bar{y}]_+ = 1.
                     \label{brackets1}
                     \end{equation}
In holomorphic representation they read \cite{Berezin,Faddeev}
        \begin{equation}
        \bar{x} \rightarrow \frac{\partial}{\partial x},
        \bar{y} \rightarrow \frac{\partial}{\partial y}.  \label{holomorphic}
        \end{equation}
With this choice, the wave function is a function of the background
variables and the unbarred variables. Then there is the usual
operator-ordering ambiguity in going from the classical constraint
(\ref{constraint}) to the Wheeler-DeWitt equation (\ref{Wheeler-DeWitt}).
Following \cite{Hall-ferm,Claus-ferm}, we shall adopt the Weyl ordering,
which in this case involves the substitution
                    \begin{equation}
                    x\bar{x} \rightarrow \frac12
                    \left(x\frac{\partial}{\partial x} -
                    \frac{\partial}{\partial x}x\right).
                    \label{Weyl}
                    \end{equation}

With the above choice of operator ordering, the Hamiltonian
(\ref{Hamiltonian}) becomes the operator
                    \begin{equation}
                    \hat H_n = -\nu + \nu \left(
                    x\frac{\partial}{\partial x}+
                    y\frac{\partial}{\partial y}\right)
                    +m\left(yx + \frac{\partial^2}
                    {\partial x \partial y}\right).
                    \label{Hamiltonian1}
                    \end{equation}
It is well known \cite{Hall-Hawk,Laflamme,Barv,we-decoh} that in the
semiclassical approximation the wave function of the Universe can be
represented as a product of the wave function of macroscopical background
variables (scale factor, inflaton, etc.)\footnote {We work in the framework
of the reduced phase space quantisation in which the wave function
explicitly depends on time. } and an infinite number of partial wave
functions describing the perturbations of microscopic -- in our case
fermionic -- fields,
                    \begin{equation}
                    \Psi(t,a,\varphi|x,y) = \Psi_0(t,a,\varphi)\,
                    \prod_n \psi_n(t|x,y)\ .
                    \label{product}
                    \end{equation}
These partial wave functions satisfy Schr\"{o}dinger equations with respect to
the semiclassical time parameter,
                    \begin{equation}
                    i\dot{\psi_n}(t|x,y) = \hat H_n \psi_n(t|x,y).
                    \label{Schrod}
                    \end{equation}
Before solving (\ref{Schrod}) with the Hamiltonian
 (\ref{Hamiltonian1}), one should introduce the inner product for
 fermionic
 wave functions appropriate to the holomorphic representation, defined for
 any pair of functions $f,g$ of the Grassmann variables $x,y$ \cite
 {Berezin,Faddeev}:
                    \begin{equation}
                    (f,g) = \int \overline{f(x,y)}g(x,y)
                    e^{-x\bar{x}-y\bar{y}}dxd\bar{x}dyd\bar{y}.
                    \label{inner-product}
                    \end{equation}
Integration over $x,y,\bar{x},\bar{y}$ is performed according to usual
rules of Berezin integration \cite{Berezin}
                    \begin{equation}
                    \int dx = 0,\  \int x dx = 1,\ \int d\bar{x} = 0,
                    \ \int \bar{x} d\bar{x} = 1,
                    \label{Berezin}
                    \end{equation}
and likewise for $y$ and $\bar{y}$.

We are now in a position to look for the solution of the Schr\"{o}dinger
equations (\ref{Schrod}). Taking into account the Grassmannian nature of
the variables $x$ and $y$, one can easily guess that this wave function is
a second-order polynomial in these variables. Knowing that
the no-boundary wave function of the Universe
\cite{Hartle-Hawk} is given by a Gaussian exponential, we shall look for
the solution of (\ref{Schrod}) in the following form:
                      \begin{equation}
                      \psi_n(t|x,y) = v_n +  \bar{v}_n x y.
                      \label{wavefunction}
                      \end{equation}

Substituting (\ref{wavefunction}) and (\ref{Hamiltonian1}) into
(\ref{Schrod}), one can get the following system of relations for the
functions $v_n$ and $\bar{v}_n$:
                      \begin{eqnarray}
                      &&i\dot{v}_n + \nu v_n + m \bar{v}_n = 0,\\
                      \label{relation}
                      &&i\dot{\bar v}_n - \nu \bar v_n + m v_n = 0.
                      \label{relation1}
                      \end{eqnarray}
These equations obviously coincide with the Dirac equations
(\ref{Dirac-eq3}) and (\ref{Dirac-eq2}) for $y$ and $\bar{x}$.
Now the wave function (\ref{wavefunction}) can be rewritten in terms of
the first of these two functions, the second being expressed in terms
of the first one:
        \begin{equation}
        \psi_n(t|x,y) = v_n-\frac{i \dot{v}_n + \nu v_n }{m}x y\ .
        \label{wavefunction1}
        \end{equation}
Eliminating $\bar v_n$ allows one to write a second-order equation for
the basis function $v_n$,
        \begin{equation}
        \ddot{v}_n + ( - i \dot{\nu}  + m^2 + \nu^2) v_n = 0, \label{second-order}
        \end{equation}
which obviously has two independent solutions. The choice of a particular
solution can be done on the basis of the path-integral derivation of the
no-boundary wave function which automatically picks out a particular vacuum
state for quantum fields on the DeSitter background. This will be done in
the next section.

\section{Path integral, Euclidean
effective action and density matrix of fermions}
\hspace{\parindent}
The definition of the no-boundary wave function of fermionic fields in the
holomorphic representation of the previous section can be given by the
path integral over the fermionic fields in the spacetime ball
with the Euclidean spacetime metric
        \begin{equation}
        ds^2_E = d\tau^2 + a^2(\tau) d\Omega_3^2,\ \
        a(\tau) = \frac{1}{H}\sin H\tau,               \label{Euclid}
        \end{equation}
which can be obtained from the Lorentzian metric by the analytic
continuation in the complex plane of time, $\tau=\pi/2H+it$. For the partial
wave functions this integral reads
        \begin{eqnarray}
        \psi_n^{\rm E}(\tau|x,y)=\int Dx\,D\bar x\,Dy\,D\bar y\,
        e^{-I_n[x,\bar x,y,\bar y]},
        \end{eqnarray}
where $I_n[x,\bar x,y,\bar y]$ is the Euclidean action following from
(\ref{action2}) by Wick rotation and by omitting the initial surface
boundary term which vanishes because in the Hartle-Hawking prescription
this surface shrinks to a point satisfying the requirement of regularity of
all fields. The integration runs over fields with fixed values of the {\em
unbarred} fermionic variables $(x,y)$ at the final (outer) boundary $S_F$
labelled by the Euclidean time $\tau$. The Gaussian integration gives the
answer
        \begin{eqnarray}
        &&\psi_n^{\rm E}(\tau|x,y)={\rm Det}\,\Big(-\frac{d^2}{d\tau^2}+
        \nu'+\nu^2+m^2\Big)\,e^{-I_n^{\rm ext}}, \nonumber\\
        &&I_n^{\rm ext}=\frac12\left[\bar x(\tau)x(\tau)
        +\bar y(\tau)y(\tau)\right],                        \label{psi}
        \end{eqnarray}
where the prime denotes derivation with respect to Euclidean time, and
$I_n^{\rm ext}$ is the value of the Euclidean action at the classical
extremal, and the functional determinant represents the one-loop
contribution of quantum fluctuations near this extremal.

The classical Euclidean equations of motion easily follow from their
Lorentzian counterpart
        \begin{eqnarray}
        &&-x' + \nu x - m\bar{y} = 0, \\
        &&-\bar x' - \nu \bar{x} + my = 0,\\
        &&-y' + \nu y + m\bar{x} = 0,\\
        &&-\bar y' - \nu \bar{y} - mx = 0.
        \end{eqnarray}
The solution of this system of equations in the Euclidean ball
$0<\tilde\tau<\tau$ with fixed values $x(\tau)=x,\,y(\tau)=y$ at the
boundary,
        \begin{eqnarray}
        &&x(\tilde\tau)=\frac{u(\tilde\tau)}{u(\tau)}x,\,\,\,
        y(\tilde\tau)=\frac{u(\tilde\tau)}{u(\tau)}y,  \nonumber\\
        &&\bar x(\tilde\tau)=
        \frac{u'(\tilde\tau)-\nu u(\tilde\tau)}{mu(\tau)}y,\,\,\,
        \bar y(\tilde\tau)=
        \frac{-u'(\tilde\tau)+\nu u(\tilde\tau)}{mu(\tau)}x,
        \end{eqnarray}
can be given in terms of one function $u(\tau)$ satisfying the Euclidean
version of (\ref{second-order}),
        \begin{eqnarray}
        u'' + (\nu' + m^2 + \nu^2) u = 0,\,\,\, u(0)={\rm reg}, \label{u}
        \end{eqnarray}
and the condition of regularity at $\tau=0$. Thus, with this solution the
exponential of the Euclidean wave function reads $-I_n^{\rm
ext}=xy(u'(\tau)-\nu u(\tau))/mu(\tau)$. The preexponential functional
determinant can also be expressed in terms of this Euclidean basis function
according to the technique of \cite{norm,reduc,reduc2,we-decoh}. As shown
in \cite{we-annals,we-decoh} this determinant, when written as a product of
eigenvalues $\lambda$ of the differential operator, can be identically
transformed as
        \begin{eqnarray}
        {\rm Det}\Big(-\frac{d^2}{d\tau^2}+
        \nu'+\nu^2+m^2\Big)=\prod_\lambda \lambda=E(0),
        \end{eqnarray}
where $E(0)$ is a particular value at $\lambda=0$ of the left-hand side of
the equation
        \begin{eqnarray}
        E(\lambda)=0
        \end{eqnarray}
for the eigenvalues of the operator in question. In our case we have a
Dirichlet problem (zero-boundary condition at the boundary of the Euclidean
ball $\tau$) for the second order differential operator with the
$\lambda=0$ eigenfunction $u(\tau)$. Therefore, $E(0)=u(\tau)$ in the
equations above, so that the needed functional determinant coincides with
the basis function taken at the boundary of the Euclidean ball,
        \begin{eqnarray}
        {\rm Det}\Big(-\frac{d^2}{d\tau^2}+
        \nu'+\nu^2+m^2\Big)=u(\tau).
        \end{eqnarray}

The resulting Euclidean no-boundary wavefunction of fermions reads
        \begin{eqnarray}
        \psi_n^{\rm E}(\tau|x,y)=u\,\exp
        \left(\frac{u'-\nu u}{mu}xy\right)
        =u+(u'-\nu u)\frac{xy}{m},                              \label{Lor-w}
        \end{eqnarray}
and, when analytically continued to the complex plane of time
$\tau=\pi/2H+it$, gives exactly the wave function (\ref{wavefunction1}) that was
obtained above by directly solving the Schr\"odinger equation. The
corresponding Lorentzian basis functions
        \begin{eqnarray}
        v(t)=u(\pi/2H+it),\,\,\,v^*(t)=u(\pi/2H-it)      \label{contin}
        \end{eqnarray}
turn out to be, respectively, the negative and positive frequency basis
function of the DeSitter invariant vacuum of fermions \cite{Kam-Mish}. This
concludes the justification of the Euclidean DeSitter invariant vacuum from
the no-boundary path integral for fermions. Such a justification in the
tree-level approximation (disregarding the preexponential factors) was achieved
for bosons in \cite{Laflamme} and for fermions in \cite{Hall-ferm}. Here it
is now proven up to the one-loop approximation. Note that for the case of
fermions, this one-loop contribution brings a nontrivial overall
time-dependent factor into the wave function (\ref{Lor-w}) which was not
correctly considered in \cite{Hall-ferm}\footnote {In \cite{Hall-ferm} the
fermionic Hamiltonian was supplied by a {\em complex} vacuum energy term,
allegedly following from the full Wheeler-DeWitt equation. This certainly
leads to nonconservation of the norm of the vacuum state differing from
(\ref{wavefunction}) by a time-dependent multiplier. Rigorous derivation of
the one-loop cosmological wave function, made for bosons in
\cite{norm,qsi,tunnel}, cannot lead to such violations of unitarity.}.

Using Grassmann integration rules (\ref{Berezin})
one can find the positive definite square of the norm of the wave function
(\ref{wavefunction1}):
        \begin{equation}
        N^2_n\equiv(\psi_n,\psi_n) =\frac1{m^2}\left(
        |\nu v_n +i \dot{v}_n|^2
        + m^2|v_n|^2\right).                 \label{norm}
        \end{equation}
Rewritten in terms of barred and unbarred components of the basis function,
$N^2=|v_n(t)|^2+|\bar v_n(t)|^2$, $\bar v_n(t)\equiv \bar u_n(\pi/2H+it)$,
this norm obviously features the conservation in time as a consequence of the
Dirac equations of motion in Lorentzian spacetime.

Let us now go over to the reduced density matrix of the Universe induced by
fermions. Integration over fermionic variables as environmental degrees of
freedom leads for the full cosmological wave function (\ref{product}) to
the expression
        \begin{eqnarray}
        \rho(a,\varphi|a',\varphi')=\Psi_0\Psi_0'^*\left(\prod_n N_n\right)\,
        \left(\prod_n N_n'\right)
        \mbox{\boldmath$D$}(a,\varphi|a',\varphi'),
        \end{eqnarray}
where the nontrivial decoherence effects due to fermions are contained in
the infinite products of norms and the decoherence factor
        \begin{eqnarray}
        &&\mbox{\boldmath$D$}(a,\varphi|a',\varphi')=
        \prod_n D_n(a,\varphi|a',\varphi'),    \\
        &&D_n(a,\varphi|a',\varphi') = \frac{v_n v_n'^*}{m^2\,N_n \,N'_n}
        \left[\,m^2 + \nu\nu'
        + \left(\frac d{dt}\ln v_n\right)
        \left(\frac d{dt}\ln v_n'^*\right)\right.           \nonumber\\
        &&\qquad\qquad\qquad\left.
        - i\nu\left(\frac d{dt}\ln v_n'^*\right)
        + i\nu'\left(\frac d{dt}\ln v_n\right)\right],    \label{decoh}
        \end{eqnarray}
where the primes obviously denote that the corresponding quantities are
calculated on the background of primed macroscopic variables $(a',\varphi')$.

The fermionic contribution to diagonal elements of the density matrix
reduces to the products of the norms of the partial wave functions. Let us
show that, as in the bosonic case \cite{norm,tunnel,qsi,we-decoh}, this contribution
is given by the Euclidean effective action of fermion fields calculated on
the spherical DeSitter instanton:
        \begin{eqnarray}
        &&\prod_n N_n=
        \exp\left(-\frac12\mbox{\boldmath$\SGamma$}\right), \label{efac1}\\
        &&\exp(-\mbox{\boldmath$\SGamma$})=
        \prod_n\int Dx\,D\bar x\,Dy\,D\bar y\,
        e^{-I_n[x,\bar x,y,\bar y]}.                         \label{efac2}
        \end{eqnarray}
Here $I_n[x,\bar x,y,\bar y]$ is again the Euclidean action, but now it is
taken on the whole sphere of the DeSitter instanton $0<\tau<\pi/H$ on the
class of fields regular at both poles $\tau=0$ and $\tau=\pi/H$. Due to the
properties of spatial harmonics, this regularity implies that the variables
simply vanish there, $(x,y)(\tau_\pm)=0,\,\tau_\pm\equiv 0,\pi/H$;  that
is why both of the surface terms corresponding to these poles are also
vanishing. The partial contribution to the effective action is obviously
given by the functional determinant of the same operator as above,
        \begin{eqnarray}
        \exp(-\mbox{\boldmath$\SGamma$}_n)=
        {\rm Det}\Big(-\frac{d^2}{d\tau^2}+
        \nu'+\nu^2+m^2\Big)=\prod_\lambda \lambda=E_n(0),
        \end{eqnarray}
but now with the other boundary conditions reflecting the regularity of
fields at $\tau_\pm$. Therefore, the left-hand side of the equation for the
zero eigenvalue $E(0)$ is here given by a qualitatively different
expression. Let us derive it and, in this way, demonstrate that it is given
by $E(0)=N^2_n$ -- the square of the norm of the partial vacuum wave
function with the quantum number $n$, see (\ref{norm}).

For this purpose, introduce the full Euclidean DeSitter sphere by extending
the metric (\ref{Euclid}) up to the second pole $\tau=\pi/H$ and smoothly
continue the operator $F=-d^2/d\tau^2+(m^2+\nu^2+\nu')$ to the second
hemisphere by observing that $\nu(\tau)=\nu(\pi/H-\tau)$. Then introduce
two sets of basis functions of this operator (solutions of the
corresponding homogeneous equation (\ref{u}))
        \begin{eqnarray}
        &&u''_\pm - (\nu' + m^2 + \nu^2) u_\pm = 0, \\
        &&u_+(0)={\rm reg},\,\,\,u_-(\pi/H)={\rm reg}.
        \end{eqnarray}
One certainly identifies $u_+(\tau)$ with the Euclidean basis function
$u(\tau)$ introduced above and also tries as a candidate for $u_-(\tau)$
the function $u_1(\tau)=u(\pi/H-\tau)$ used in the bosonic case in
\cite{we-decoh}. Now, however, this turns out to be impossible, because in
view of the relation $\nu'(\pi/H-\tau)=-\nu'(\tau)$ this function satisfies
another equation, $u''_1 - (-\nu'+m^2+\nu^2)u_1 = 0$. The remedy is to take
first the new function $u_2=(u'-\nu u)/m^2$ which, as can easily be
verified, satisfies this equation with the negative sign of $\nu'$, $u''_2
- (-\nu'+m^2+\nu^2)u_2 = 0$, and then construct its mirror image with
respect to the equatorial section of $\tau=\pi/2H$,
$u_-(\tau)=u_2(\pi/H-\tau)$. Then this function will satisfy the original
equation and will be, by construction, regular at the pole $\tau=\pi/H$.
Thus the set of two basis functions of our operator regular at the opposite
poles of the DeSitter instanton is given by
        \begin{eqnarray}
        u_+(\tau)=u(\tau),\,\,\,u_-(\tau)=\left.\frac{u'-\nu
        u}{m^2}\right|_{\tau=\pi/H-\tau}.
        \end{eqnarray}
In terms of these two basis functions, the equation of the zero eigenvalue
of the operator can be written as the linear dependence of two
two-dimensional vectors composed of these functions and their derivatives,
        \begin{eqnarray}
        \left[\,\begin{array}{c}
        u_+(\tau)\\
        u_+'(\tau)
        \end{array}\,\right]-\sigma\left[\,\begin{array}{c}
        u_-(\tau)\\
        u_-'(\tau)
        \end{array}\,\right]=0.
        \end{eqnarray}
The left-hand side of this equation can also be written as
        \begin{eqnarray}
        E(0)=\Big(u_-u'_+-u_+u'_-\Big)(\tau).
        \end{eqnarray}
Note that in view of the Wronskian relation for the operator in question,
this quantity is independent of $\tau$. Therefore, when calculated at the
equator $\tau=\pi/2H$, on using the equation for $u(\tau)$ and its
corollary $u'_-(\tau)=(1/m^2)(-mu+(u'-\nu u)\nu)(\pi/H-\tau)$, the
quantity $E(0)$ reads
        \begin{eqnarray}
        E(0)=\frac1{m^2}\left[\,m^2u^2+(u'-\nu u)^2\,\right]_{\tau=\pi/2H}.
        \end{eqnarray}
In view of the analytic continuation to the Lorentzian time
$\tau=\pi/2H+it$, (\ref{contin}) at $\tau=\pi/2H$ -- the point of
``nucleation'' of the Lorentzian spacetime from the Euclidean hemisphere --
we have $u(\pi/2H)=v(0)=v^*(0)$, $u'(\pi/2H)=-i\dot v(0)=i\dot v^*(0)$ and
        \begin{eqnarray}
        E(0)=N^2_n
        \end{eqnarray}
with the square of the norm given by the expression (\ref{norm}) -- the
quantity which is in its turn time-independent in view of the Dirac equation
for basis functions in Lorentzian spacetime.

This finally proves the relations (\ref{efac1})-(\ref{efac2}) and brings the
fermionic contribution to the reduced density matrix in quantum cosmology to the
form
        \begin{eqnarray}
        \rho_{\rm fermion}(a,\varphi|a'\varphi')=
        \exp\left(-\frac12\mbox{\boldmath$\SGamma$}
        -\frac12\mbox{\boldmath$\SGamma$}'\right)
        \mbox{\boldmath$D$}(a,\varphi|a',\varphi'),
        \end{eqnarray}
analogous to the bosonic contribution obtained in our previous paper
\cite{we-decoh}.

This concludes the proof of the Euclidean effective-action algorithm for
the diagonal element of the density matrix -- the distribution function of
inflationary cosmological models -- derived in the bosonic case in the
series of publications \cite{norm,tunnel,qsi}, and now extended to
fermions. For the off-diagonal elements the effective action factors
$\exp(-\frac12\mbox{\boldmath$\SGamma$})$ and
$\exp(-\frac12\mbox{\boldmath$\SGamma$}')$ also modify the usual
expression \cite{Claus-ferm}, but these factors are constant in $t$ and,
 therefore, become
negligible for late Lorentzian times, provided the decoherence effects due
to $\mbox{\boldmath$D$}(a,\varphi|a',\varphi')$ rapidly grow with time.
Therefore, in what follows we shall concentrate on the time evolution of
the decoherence factor $\mbox{\boldmath$D$}(a,\varphi|a',\varphi')$.

\section{Decoherence due to massless versus massive fermions}
\hspace{\parindent}
The structure of the decoherence factor for fermions is much more
complicated than the corresponding one for bosons (cf. \cite{we-decoh}).
 Their quantum state
(\ref{wavefunction1}) generated according to the no-boundary prescription
can be interpreted as  the vacuum state corresponding to the following
creation and annihilation operators:
                      \begin{eqnarray}
                      &&a_n = \frac{|\nu v_n -i \dot{v}_n|}
                      {\sqrt{|\nu v_n -i \dot{v}_n|^2 + m^2|v_n|^2}}
                      \frac{\partial}{\partial y}
                      +\frac{|v_n|m}
                      {\sqrt{|\nu v_n -i \dot{v}_n|^2 + m^2|v_n|^2}} x,
                      \nonumber \\
                      &&a_n^+ = \frac{|\nu v_n -i \dot{v}_n|}
                      {\sqrt{|\nu v_n -i \dot{v}_n|^2 + m^2|v_n|^2}}y
                      +\frac{|v_n|m}
                      {\sqrt{|\nu v_n -i \dot{v}_n|^2 + m^2|v_n|^2}}
                      \frac{\partial}{\partial x},
                      \nonumber \\
                      &&b_n = \frac{|\nu v_n -i \dot{v}_n|}
                      {\sqrt{|\nu v_n -i \dot{v}_n|^2 + m^2|v_n|^2}}
                      \frac{\partial}{\partial x}
                      -\frac{|v_n|m}
                      {\sqrt{|\nu v_n -i \dot{v}_n|^2 + m^2|v_n|^2}} y,
                      \nonumber \\
                      &&b_n^+ = \frac{|\nu v_n -i \dot{v}_n|}
                      {\sqrt{|\nu v_n -i \dot{v}_n|^2 + m^2|v_n|^2}}x
                      -\frac{|v_n|m}
                      {\sqrt{|\nu v_n -i \dot{v}_n|^2 + m^2|v_n|^2}}
                      \frac{\partial}{\partial y}.           \label{crea-annih}
                      \end{eqnarray}

In the {\em massless} limit $m = 0$, these operators have a very simple form:
                     \begin{eqnarray}
                     &&a_n = \frac{\partial}{\partial y},\nonumber \\
                     &&b_n = \frac{\partial}{\partial x},
                     \label{crea-annih1}
                     \end{eqnarray}
and the vacuum state contains only the Grassmann-independent component
following from the Schr\"{o}dinger equation (\ref{Schrod}):
                    \begin{equation}
                    \psi_n = \exp\Big(i\int \nu(t) dt\Big).
                    \label{massless}
                    \end{equation}
{}From the structure of the wave function (\ref{massless}) one can see that
the decoherence factor does not have a real part and hence, decoherence
effects are absent. This phenomenon was already described in
\cite{Claus-ferm}. It is not surprising because the absence of decoherence
was observed not only for conformally invariant massless fermions but for
all conformally invariant fields such as the electromagnetic field and
conformally-coupled massless scalar field \cite{we-decoh}.

Let us turn now to the case of a {\em massive} spinor field. To get the
Hartle-Hawking vacuum one should take the wave function
(\ref{wavefunction1}), where the function $v_n$ can be obtained by analytic
continuation from the solution of the Euclidean counterpart of the second-order
equation (\ref{second-order}). One should choose a solution of the
Euclidean equation which is regular at the Euclidean hemisphere and serves
as a germ for the Lorentzian Universe \cite{tunnel,qsi}. Such a solution
can be found for example in  \cite{Kam-Mish} and has the following form:
                \begin{eqnarray}
                &&v_n = (1+i\sinh Ht)^{n/2+3/4}(1-i\sinh Ht)^{-n/2+1/4}
                \nonumber \\
                &&\times _2 F _1 (1 + im/H, 1 -im/H;n+2; (1 + i\sinh Ht)/2).
                \label{basis-function}
                \end{eqnarray}
We are interested in the case of large masses because at the beginning of
inflation these large masses can be generated by Yukawa interactions
between fermions and the inflaton scalar field (see, for example
\cite{norm,qsi}). In this case it would be convenient to have a uniform
asymptotic expansion for the solution of (\ref{second-order}).
Unfortunately, in contrast to Legendre or Bessel equations (cf.
\cite{Thorne,we-annals}), such an expansion cannot be written down in
closed form. However, one can write down the main terms of this expansions
responsible for the leading structure in the decoherence factor
(\ref{decoh}). The leading-in-mass $m$ part of the basis-function frequency
looks like
          \begin{equation}
          \frac d{dt}\ln v_n \sim
          \frac ia \sqrt{n^2 + m^2a^2}.  \label{basis-function1}
          \end{equation}
Substituting it into the formula for the decoherence factor $D_n$, one gets
the following expression for the leading term of the amplitude for the
total decoherence factor:
        \begin{eqnarray}
        &&|\mbox{\boldmath$D$}(a,\varphi|a',\varphi')| = \prod_n |D_n| =
        \exp \left(-\frac{m^2(a-a')^2}8\,I\right),    \label{decoh1}\\
        &&I = \sum_{n=1}^{\infty}
        \frac{n(n+1)}{(n+1/2)^2}.                    \label{decoh2}
        \end{eqnarray}
It is interesting to notice that in the opposite limit of small masses $m$
one can immediately use the hypergeometrical expansion
(\ref{basis-function}), and the leading term in mass again has the
structure (\ref{decoh1}). This expression for small mass was obtained
earlier in \cite{Claus-ferm}.

It is clear that the expression (\ref{decoh1}) contains linear ultraviolet
divergences. It is in sharp contrast with the case of boson fields, studied
in our preceding paper \cite{we-decoh}. Indeed, in the case of boson
fields, using conformal parametrisation of fields, it was possible to avoid
ultraviolet divergences (see also \cite{Claus-elec}). Now, in spite of
using the conformal parametrisation, this is not the case here. Moreover, we cannot
freely redefine the fermion field by multiplying it with a power of the
cosmological factor $a$, because the measure on the corresponding Hilbert
space of states (see (\ref{inner-product})) then starts depending on this
multiplier. With two scale factors $a$ and $a'$ related to the arguments of
the density matrix this would mean having two different integration
measures, neither of which can be used for constructing the reduced density
matrix. Thus, we should try to renormalise the decoherence factor by some
regularisation procedure. This will be done in the next section by means of
dimensional regularisation \cite{dimen,Bar-Vil}.

\section{Renormalised fermionic decoherence factor}
\hspace{\parindent}
We would like to calculate now the renormalised expression for the
decoherence factor (\ref{decoh1}). This calculation is reduced to the
calculation of the regularised sum (\ref{decoh2}). If this sum  is
positive, then there is a decoherence effect, while a negative sign of
this quantity would imply intrinsic inconsistency of the obtained density
matrix which is likely to become an unbounded operator \cite{we-decoh}.

In the framework of dimensional regularisation, ${\rm dim}(n,4) = n(n+1)$
should be substituted by ${\rm dim}(n,d)$ from (\ref{dimen1}), and the
eigenvalue of the three-dimensional projection of the Dirac operator,
$(n+1/2)$, should be substituted by that of the $(d-1)$ - dimensional Dirac
operator, $(n + (d-3)/2)$. Thus,
 one has
           \begin{equation}
           I = \sum_{n=1}^{\infty} \frac{{\rm dim}(n,4)}{(n+(d-1)/2)^2}.
           \label{decoh3}
           \end{equation}

To calculate the sum of this type, one can use the Euler-Maclaurin formula:
           \begin{eqnarray}
           &&I = I_1 + I_2,  \label{decoh4}\\
           &&I_1 = \int_0^{\infty} dy
           \frac{{\rm dim}(y,d)}{(y+(d-3)/2)^2},  \label{I1-def}\\
           &&I_2 = i\int_0^{\infty}\frac{dy}{\exp(2\pi y) - 1}
           \left(\frac{{\rm dim}(iy,d)}{(iy+(d-3)/2)^2} -
           \frac{{\rm dim}(-iy,d)}{(-iy+(d-3)/2)^2}\right). \label{I2-def}
           \end{eqnarray}
The last integral (see (\ref{I2-def})) is convergent at $d = 4$ and equals
\cite{Grad-Ryz}:
           \begin{equation}
           I_2 = \frac12\int_0^{\infty}\frac{dy y}
           {(\exp(2\pi y) - 1)(y^2 + 1/4)^2} =
           \frac{\pi^2}{8} - 1.                           \label{I2-def1}
           \end{equation}

The integral $I_1$ in  (\ref{I1-def}) is linearly divergent and for its
regularisation it is necessary to use the analytic-continuation procedure
and integration by parts typical for dimensional regularisation
\cite{Bar-Vil,we-decoh}. Now it is convenient to make a change of variables
from $y$ to $x = 1/y$ and to introduce the new quantity
           \begin{equation}
           f(x,d) = \frac{4 {\rm dim} (1/x, d)}{[x(d-3) + 2]^2}
           x^{2^{d/2} - 2},
           \label{f-define}
           \end{equation}
which is finite at $x \rightarrow \infty$.
Now, the integral $I_1$ can be rewritten as follows:
           \begin{equation}
           I_1 = \int_0^{\infty} dx f(x,d) x^{-(2^{d/2} - 2)}.
           \label{I1-def1}
           \end{equation}
Integrating two times by part and discarding the terms corresponding to
power-law divergences, one gets
           \begin{equation}
           I_1 = \frac{1}{(2^{d/2} - 3)(2^{d/2} - 4)}
           \int_0^{\infty} dx f''(x,d)  x^{-(2^{d/2} - 4)}.
           \label{I1-def2}
           \end{equation}
Then, expanding expression (\ref{I1-def3}) in powers of $(d-4)$ one has
the following regularised expression:
           \begin{equation}
           I_1 = -\int_0^{\infty} dx \ln x f''(x,4) -
           \frac{f'(0,4)}{2^{d/2}-4} + f'(0,4)
           -\frac{f'_{,d}(0,4)}{2\ln 2},
           \label{I1-def3}
           \end{equation}
where $,d$ denotes the derivative with respect to the dimensionality of
spacetime. The direct calculation shows that
           \begin{equation}
           f'(0,4) = 0,
           \label{I1-def4}
           \end{equation}
which corresponds to the absence of logarithmic ultraviolet divergences.
Then
           \begin{equation}
           \int_0^{\infty} dx \ln x f''(x,4) = + \frac12,
           \label{I1-def5}
           \end{equation}
and the mixed derivative is given by
           \begin{equation}
           f'_{,d}(0,4) = -1 + 2 \ln 2.
           \label{mixed}
           \end{equation}
Substituting (\ref{I1-def4})--(\ref{mixed}) into (\ref{I1-def3}), one has
           \begin{equation}
           I_1 = \frac{1 - 3 \ln 2}{2},
           \label{I1}
           \end{equation}
and combining (\ref{I1}) with (\ref{I2-def1}) one finally has
           \begin{equation}
           I = \frac{1 - 5 \ln 2}{2 \ln 2} +
           \frac{\pi^2}{8} < 0.
           \label{negative}
           \end{equation}
Thus, we have found a negative result which would be in strong
contradiction with basic properties of a
density matrix, since ${\rm Tr}\rho^2$ would diverge.
The next section will show how the divergences can be remedied without
leading to inconsistencies.

\section{Bogoliubov transformation for fermions and decoherence}
\hspace{\parindent}
We have already seen in Sec. 5 that ultraviolet divergences arise in the
decoherence factor of the reduced density matrix coming from fermionic
perturbations, and in Sec. 6 we have obtained a dimensionally-renormalised
expression which manifested wrong, anti-decoherent behaviour, breaking
such fundamental properties of the density matrix as the normalisabity of its
square which is equivalent to its positive-definiteness and boundedness. It
happens in spite of working from the beginning in the conformal
parametrisation for the spinor field, which was efficient in eliminating
ultraviolet divergences in decoherence factors induced by bosonic
perturbations \cite{Claus-elec,we-decoh}. Moreover, we do not have the
possibility to change the parametrisation of fermion fields by simply
multiplying them by some power of the scale factor $a$, as it was done for
bosons \cite{we-decoh}. This happens because, as was mentioned in Sec. 5, the inner
product of wave functions depending on fermion fields depends on these
fields.

However, we still have some additional freedom of redefining the
environmental variables by performing
some kind of Bogoliubov transformations. This will lead to
well-defined expressions.
At the end of this section it will be shown that this
transformation is analogous to the well-known Foldy-Wouthuysen
transformation \cite{Itzyk-Zub}.

Thus, let us make the following Bogoliubov transformation
        \begin{eqnarray}
        &&x = \alpha \tilde{x} + \beta\bar{\tilde{y}},
        \nonumber \\
        &&y = \gamma\tilde{y} + \delta \bar{\tilde{x}},
        \nonumber \\
        &&\bar{x} = \alpha^* \bar{\tilde{x}} + \beta^*\tilde{y},
        \nonumber \\
        && \bar{y} = \gamma^*\bar{\tilde{y}} + \delta^* \tilde{x}.
        \label{Bogol}
        \end{eqnarray}
Substituting the expressions (\ref{Bogol}) into the action (\ref{action2})
and requiring the conservation of the form of the kinetic term
$\bar{x}\dot{x} + \bar{y}\dot{y}$, one has the following requirements for
the Bogoliubov coefficients $\alpha,\beta,\gamma,\delta$ :
        \begin{eqnarray}
        &&\alpha^* \beta + \gamma^* \delta = 0, \nonumber \\
        &&|\alpha|^2 + |\beta|^2 = 1, \nonumber \\
        &&|\gamma|^2 + |\delta|^2 = 1.
        \label{Bogol1}
        \end{eqnarray}
It is convenient to choose the Bogoliubov coefficients in the form:
        \begin{eqnarray}
        &&\alpha = \gamma, \nonumber \\
        &&\beta = - \delta.
        \label{Bogol2}
        \end{eqnarray}

Now, substituting the transformed Grassmann variables (\ref{Bogol}) into
the other terms of the action (\ref{action2}), taking into account
(\ref{Bogol1}) and (\ref{Bogol2}) and arriving at the Hamiltonian
formalism, one can get the new Hamiltonian
        \begin{equation}
         \tilde{H}_n(\tilde{x},\bar{\tilde{x}},\tilde{y},\bar{\tilde{y}})
          = \tilde{\nu}(\tilde{x}\bar{\tilde{x}}+\tilde{y}\bar{\tilde{y}})
         + \tilde{m}^*\tilde{y}\tilde{x} + \tilde{m}\bar{\tilde{x}}
         \bar{\tilde{y}},
         \label{Hamiltonian2}
         \end{equation}
where
        \begin{eqnarray}
        &&\tilde{\nu} = i (\dot{\alpha}\alpha^* + \beta\dot{\beta}^*)
        + \nu (|\alpha|^2 - |\beta|^2) - m (\alpha\beta +
        \alpha^*\beta^*), \ \       \tilde{\nu}^* = \tilde{\nu}\\
        \label{nu-new}
        &&\tilde{m} = m(\alpha^{*2} - \beta^2) - 2\nu\alpha^*\beta
        + i(\beta\dot{\alpha}^* - \dot{\beta}\alpha^*).
        \label{mass-new}
        \end{eqnarray}

Thus, we have found a new Hamiltonian which has the same form as the old
one (\ref{Hamiltonian}), with the only difference that $\nu$ and $m$ are
replaced by new functions $\tilde{\nu}$ and $\tilde{m}$ depending on
Bogoliubov coefficients. Besides, the ``mass'' $\tilde{m}$ is complex now
($\tilde\nu$ remains real as one can check by using the relation
(\ref{Bogol1}) for Bogoliubov coefficients).

The Dirac equations now read
        \begin{eqnarray}
        &&i\dot{\tilde{x}} + \tilde{\nu} \tilde{x} -
        \tilde{m}\bar{\tilde{y}} = 0,\\
        \label{Dirac-eq1n}
        &&i\dot{\bar{\tilde{x}}} - \tilde{\nu}
        \bar{\tilde{x}} + \tilde{m}^*\tilde{y} = 0,\\
        \label{Dirac-eq2n}
        &&i\dot{\tilde{y}} + \tilde{\nu} \tilde{y} +
        \tilde{m}\bar{\tilde{x}} = 0,\\
        \label{Dirac-eq3n}
        &&i\dot{\bar{\tilde{y}}} - \tilde{\nu}\bar{\tilde{y}}
         - \tilde{m}^*\tilde{x} = 0.
         \label{Dirac-eq4n}
         \end{eqnarray}

For generic Bogoliubov coefficients $(\alpha_n,\beta_n)$ depending on the
quantum number $n$, the decoherence factor (\ref{decoh}) becomes
       \begin{eqnarray}
        &&\mbox{\boldmath$D$}(a,\varphi|a'\varphi')=
        \prod_n D_n(a,\varphi|a',\varphi'),    \\
        &&D_n(a,\varphi|a'\varphi') = \frac{v_n v_n'^*}{\tilde{m}_n
        \tilde{m'}^*_n\,N_n \,N'_n}
        \left[\,\tilde{m}_n\tilde{m}^*_n + \tilde{\nu}\tilde{\nu}'
        + \left(\frac d{dt}\ln v_n\right)
        \left(\frac d{dt}\ln v_n'^*\right)\right.           \nonumber\\
        &&\qquad\qquad\qquad\left.
        - i\tilde{\nu}\left(\frac d{dt}\ln v_n'^*\right)
        + i\tilde{\nu}'\left(\frac d{dt}\ln v_n\right)\right],    \label{decoh-n}
        \end{eqnarray}
with the new mass parameter $\tilde{m}_n$ also depending on $n$ ($\nu$ was
$n$-dependent even in the old parametrisation), while the approximate
expression for a potentially divergent part of this factor (\ref{decoh1})
becomes
        \begin{eqnarray}
        &&|\mbox{\boldmath$D$}(a,\varphi|a',\varphi')| = \prod_n |D_n| =
        \exp \left(-\frac{(a-a')^2}{8}\,\sum_{n=1}^\infty
        \frac{n(n+1)\tilde m_n\tilde m'^*_n}{(n+1/2)^2}
        \right),                                                \label{decoh1n}
        \end{eqnarray}
replacing (\ref{decoh1})-(\ref{decoh2}).

The dependence of the new functions $\tilde{\nu}$ and $\tilde{m}$ on
Bogoliubov coefficients gives new opportunities regarding the finiteness
and consistency of the decoherence factor. In particular, the modified mass
$\tilde{m}$ can be demanded to vanish. This would, however, lead to an absence
of decoherence. For this reason, a physical principles must be invoked
to select an appropriate Bogoliubov transformation: We demand that
there be no effect for a {\em stationary} spacetime, in accordance with the
idea \cite{we-decoh} that decoherence is connected with particle creation.
The decoherence factor should thus contain terms that vanish when
$\dot{a}$ vanishes. Since one the main features of decoherence is its
{\em irreversibility} \cite{decoherence1} -- and the escape of
created particles is, similar to usual scattering situations, an
irreversible process -- we think that the above principle captures
the physical effect of decoherence.

We therefore choose real $\alpha$
and $\beta$ satisfying the equation (this is the
$(\dot\alpha,\dot\beta)$-independent part of $\tilde m$)
        \begin{equation}
        m(\alpha^2_n - \beta^2_n)- 2\nu\alpha_n\beta_n=0.
        \label{algebraic}
        \end{equation}
One finds the following solutions:
        \begin{eqnarray}
        &&\alpha_n = \left(\frac{\sqrt{m^2 + \nu^2} + \nu}
        {2\sqrt{m^2 + \nu^2}}\right)^{1/2},\\
        \label{alpha}
        &&\beta_n = \left(\frac{\sqrt{m^2 + \nu^2} - \nu}
        {2\sqrt{m^2 + \nu^2}}\right)^{1/2}.
        \label{beta}
        \end{eqnarray}

If $\nu$ is independent of time (a stationary Universe), this
choice of Bogoliubov coefficients results in a vanishing $\tilde m$; however,
for a nonstationary Universe there is an imaginary contribution proportional
to time derivatives of Bogoliubov coefficients,
        \begin{eqnarray}
        &&\dot{\alpha}_n = \frac{m^2\dot{\nu}}{4(m^2+\nu^2)^{3/2}}
        \left(\frac{\sqrt{m^2 + \nu^2} + \nu}
        {2\sqrt{m^2 + \nu^2}}\right)^{-1/2},\\
        \label{alpha1}
        &&\dot{\beta}_n = - \frac{m^2\dot{\nu}}{4(m^2+\nu^2)^{3/2}}
        \left(\frac{\sqrt{m^2 + \nu^2} - \nu}
        {2\sqrt{m^2 + \nu^2}}\right)^{-1/2}.
        \label{beta1}
        \end{eqnarray}
Therefore,
        \begin{equation}
        \tilde{m}_n = i \frac{m \dot{\nu}}{2(m^2+\nu^2)},
        \label{m-new3}
        \end{equation}
and at large values of $n$ this mass parameter behaves as
        \begin{equation}
        \tilde{m}_n \sim \frac1n\ .
        \end{equation}

Then, turning to the expression (\ref{decoh1n})
 one can see
that the dependence of $\tilde{m}$ on $n$ presented in (\ref{m-new3})
provides the suppression of ultraviolet divergences in the modified
expression (\ref{decoh1n}). The terms in the sum in
the exponent of (\ref{decoh1n}) approach for large $n$ the expression
\be
 \frac{n(n+1)\tilde m_n\tilde m^*_n}{(n+1/2)^2}\to
  \frac{m^2\dot{a}^2}{4n^2}\ ,
\ee
and the sum is explicitly convergent.

We want to conclude this section by comparing the above Bogoliubov transformations with
their counterparts in a stationary flat spacetime. As it was already
mentioned for the case of stationary background, coefficients $\alpha$ and
$\beta$ from Eqs. (\ref{alpha})-(\ref{beta}) eliminate mass exactly. The
form of these coefficients reminds that of a Foldy-Wouthuysen
transformation \cite{Itzyk-Zub}, but really their role is just the
opposite. To explain it, let us remember that in the canonical basis (cf.
with (\ref{can-bas}) in Sec. 2), a Dirac spinor is represented in the form
        \begin{equation}
        \Psi_C = \left(
        \begin{array}{c}
        \phi\\
        \chi
        \end{array}\right),
        \label{spin-can}
        \end{equation}
where two-component spinors satisfy the following Dirac equations:
        \begin{eqnarray}
        &&i\frac{\partial \phi}{\partial t} = \vec{\sigma}\vec{p}\chi
        + m\phi,\nonumber \\
        &&i\frac{\partial \chi}{\partial t} = \vec{\sigma}\vec{p}\phi
        - m\chi.
        \label{Dir-eq-can}
        \end{eqnarray}
To disentangle these equations and to get rid of the so-called odd terms
proportional to $\vec{\sigma}\vec{p}$, one can use a Foldy-Wouthuysen
transformation
         \begin{equation}
         \Psi_C = e^{-S}\Psi_{FW},
         \label{Foldy}
         \end{equation}
where
        \begin{eqnarray}
        &&S = -i\beta \frac{\vec{\alpha}\vec{p}}{|\vec{p}|}\theta,
        \nonumber \\
        &&\beta = \left(\begin{array}{cc}
        I\  &\  0\\
        0\  &\  -I
        \end{array}\right),\ \
        \alpha = \left(\begin{array}{cc}
         0\  &\  \sigma\\
        \sigma\  &\  0
        \end{array}
        \right),\nonumber \\
        &&\theta = \frac12 \arcsin \frac{|\vec{p}|}{\sqrt{m^2 + \vec{p}^2}}.
        \label{Foldy1}
        \end{eqnarray}
This transformation also digonalises the Hamiltonian which is now
represented by a direct sum of two non-local Hamiltonians $\pm\sqrt{m^2 +
\vec{p}^2}$:
        \begin{equation}
        H = \beta \sqrt{m^2 + \vec{p}^2}.
        \label{Foldy2}
        \end{equation}
It is clear that these square roots cannot be represented in configuration
space by a finite set of differential operators. It is obvious also that
our modified Hamiltonians (\ref{Hamiltonian2}) are also combined into a
non-local Hamiltonian because all parameters $\tilde{\nu}, \tilde{m}$
of these partial Hamiltonians depend on the wave number $n$.

However, as a matter of fact, we are working in the Weyl representation for
Dirac spinors (\ref{Dir-define}) and $\gamma$ matrices
(\ref{gamma-define}). In this representation, especially, massive terms are
mixing in the Dirac equations (\ref{Dirac-eq}) and our task is to get rid
of them. This is exactly the problem that was studied in this section
 in two-component formalism;
it can thus be called ``anti-Foldy-Wouthuysen'' transformation and can be in
a traditional
(four-spinor) formalism presented by the following operators:
        \begin{equation}
        \Psi = e^{T}\tilde{\Psi},
        \label{anti-Foldy}
        \end{equation}
where
        \begin{eqnarray}
        &&T = \tilde{\theta}\frac{\vec{\alpha}\vec{p}}{|\vec{p}|}
        \left(\begin{array}{cc}
        0\  &\  I   \\
        -I\  &\  0\end{array}\right),\nonumber \\
        &&\tilde{\theta} = -\frac12\arcsin \frac{m}{\sqrt{m^2 + \vec{p}^2}}.
        \label{anti-Foldy1}
        \end{eqnarray}
Correspondingly, the Hamiltonian is turned by this transformation into
        \begin{equation}
        H = -\beta \frac{\vec{\alpha}\vec{p}\sqrt{m^2 + \vec{p}^2}}
        {|\vec{p}|}
        \label{anti-Foldy2}
        \end{equation}
and is again apparently non-local.

\section{Conclusions}
\hspace{\parindent}
Let us briefly recapitulate the main results of the present paper. We
developed the formalism of the one-loop no-boundary wave function of the
Universe in a cosmological model with fermions establishing the
connection between the path integral, effective action and semiclasssical
Schr\"odinger equation. It is worth adding here that these results are
valid for the tunneling wave function \cite{tun} as well due to the fact
that the structure of the microscopic perturbation part of these two wave
functions is the same \cite{Vach-Vil,we-tun}. We studied the expression for the
reduced density matrix and showed that the amplitude of
its decoherence factor reveals ultraviolet divergences, which was
previously observed in \cite{Claus-ferm}. The renormalised expression for
this decoherence factor which was obtained by dimensional regularization
has a wrong sign and violates the properties of density matrix as a bounded
operator. However, by making special Bogoliubov transformation for
fermionic variables one can get a finite expression for this decoherence
factor.

Thus, as in the case of bosonic variables \cite{we-decoh}, the definition
of the degrees of freedom constituting an environment is of crucial
importance for the construction of consistent expressions for the reduced
density matrix (see also
\cite{Claus-elec,Okamura}). However, there is an essential difference
between the fermionic and bosonic cases. In the case of bosonic degrees of
freedom we have a well-defined criterion for the choice of the demarcation
line between the system under consideration and the environment.
Unfortunately, the situation is not so clear for fermionic degrees of
freedom. We have put forward the suggestion to fix the Bogoliubov transformation
by demanding that the corresponding decoherence effect would be absent
if the spacetime were stationary.

One problem is that any choice of
this transformation breaks locality of a fermion field; this can create
difficulties at the consideration of fermions intereacting with other
fields. However, in contrast with bosons when the logarithm of a
decoherence factor is proportional to $m^3$ in the limit of large mass, for
fermions that is proportional only to $m^2$. This fact gives support
to the conclusion that bosons are much more efficient than fermions
for the quantum-to-classical transition \cite{Claus-ferm}.

In summary, we have extended the Euclidean effective action algorithm
for the quantum distribution function of cosmological models and their
density matrix to fermions; this was previously only known for bosons
\cite{norm,tunnel,qsi,we-tun}. Together with \cite{we-decoh} this gives
a complete account of the quantum-to-classical transition at the onset
of inflation -- the beginning of our classical Universe.

\section*{Acknowledgements}
\hspace{\parindent}
The work of A.B. was partially supported by RFBR under the grant No
96-01-00482. The work of A.K. was partially supported by RFBR under the
grant 96-02-16220 and under the grant for support of leading
scientific schools 96-15-96458. A.B. and A.K. kindly acknowledge
financial support by the DFG grants 436 RUS 113/333/4 during their
visit to the University of Freiburg in autumn~1998.


\begin{thebibliography}{99}
\bibitem{Hartle-Hawk}J.B. Hartle and S.W. Hawking, Phys. Rev. D 28 (1983) 2960.
\bibitem{Hawk}
S.W. Hawking, Nucl. Phys. B 239 (1984) 257.
\bibitem{tun}A.D. Linde, JETP 60 (1984) 211, Lett. Nuovo Cim. 39 (1984)
        401; V.A. Rubakov, Phys. Lett. B 148 (1984) 280; A. Vilenkin,
        Phys. Lett. B 117 (1982) 25, Phys. Rev. D 30 (1984) 549;
        Ya. B. Zeldovich and A.A. Starobinsky,
        Sov. Astron. Lett. 10 (1984) 135.
\bibitem{norm}A.O. Barvinsky and A.Yu. Kamenshchik, Class. Quantum Grav.
        7 (1990) L181.
\bibitem{tunnel}A.O. Barvinsky and A.Yu. Kamenshchik, Phys. Rev. D 50 (1994)
        5093.
\bibitem{qsi}A.O. Barvinsky and A.Yu. Kamenshchik, Phys. Lett. B 332 (1994) 270;
         Nucl. Phys. B 532 (1998) 339;
        A.O. Barvinsky, A.Yu. Kamenshchik and I.V. Mishakov, Nucl. Phys. B 491
        (1997) 387.
\bibitem{Zeh} H.D. Zeh, Found. Phys. 1 (1970) 69; in: Foundations
        of Quantum Mechanics, ed. by B.~d'Espagnat (Academic Press,
        New York, 1971); Found. Phys. 3 (1973) 109; O. K\"ubler
        and H.D. Zeh, Ann. Phys. (N.Y.) 76 (1973) 405.
\bibitem{decoherence1} D. Giulini, E. Joos, C. Kiefer, J. Kupsch,
      I.-O. Stamatescu and H.D. Zeh, Decoherence and the Appearance of
     a Classical World in Quantum Theory (Springer, Berlin, 1996).
\bibitem{decoherence2} C. Kiefer and E. Joos, in: Quantum Future,
    ed. by P. Blanchard and A. Jadczyk (Springer, Berlin, 1999).
\bibitem{Haroche} M. Brune, E. Hagley, J. Dreyer, X. Ma\^{\i}tre, A. Maali,
     C. Wunderlich, J.M. Raimond and S. Haroche, Phys. Rev. Lett.
     77 (1996) 4887.
\bibitem{Zeh2} H.D. Zeh, Phys. Lett. A 116 (1986) 9.
\bibitem{Kiefer1} C. Kiefer, Class. Quantum Grav. 4 (1987) 1369.
\bibitem{Polarski} C. Kiefer, D. Polarski and A.A. Starobinsky,
    Int. J. Mod. Phys. D 7 (1998) 455; C. Kiefer and D. Polarski,
    Ann. Phys. (Leipzig) 7 (1998) 137; C. Kiefer, J. Lesgourgues,
    D. Polarski and A.A. Starobinsky, Class. Quantum Grav. 15 (1998) L67.
\bibitem{Paz}J.P. Paz and S. Sinha, Phys. Rev. D 45 (1992) 2823.
\bibitem{we-decoh}A.O. Barvinsky, A.Yu. Kamenshchik, C. Kiefer and I.V. Mishakov,
        Report gr-qc/9812043, submitted to Nucl. Phys. B.
\bibitem{Claus-elec}
C. Kiefer, Phys. Rev. D 46 (1992) 1658.
\bibitem{Okamura}T. Okamura, Prog. Theor. Phys. 95 (1996) 95.
\bibitem{Claus-ferm}
C. Kiefer, Class. Quantum Grav. 6 (1989) 561.
\bibitem{dimen}
G. 't Hooft and M. Veltman, Nucl. Phys. B 44 (1972) 189;
50 (1972) 318;
C.G. Bollini and J.J. Giambiagi, Nuovo Cim. B 12 (1972) 20.
\bibitem{Bar-Vil}
A.O. Barvinsky and G.A. Vilkovisky, Phys. Rep. 119 (1985) 1.
\bibitem{Itzyk-Zub}
C. Itzykson and J.-B. Zuber, Quantum field theory (McGraw-Hill, New York,
1980).
\bibitem{Hall-ferm}
P.D. D'Eath and J.J. Halliwell, Phys. Rev. D 35 (1987) 1100.
\bibitem{we-Kluw}
G. Esposito, A.Yu. Kamenshchik and G. Pollifrone, Euclidean quantum gravity
on manifolds with boundary (Fundamental Theories of Physics, v. 85, Kluwer
Academic, Dordrecht, 1997).
\bibitem{Esp-Camb}
G. Esposito, Dirac operator and spectral geometry (Cambridge Lecture Notes
in Physics, v. 12, Cambridge Univ. Press, Cambridge, England, 1998).
\bibitem{Wess}
J. Wess and J. Bagger, Supersymmetry and supergravity (Princeton
University Press, Princeton, 1983).
\bibitem{Bog-Shir}
N.N. Bogoliubov and D.V. Shirkov, Introduction to the theory of quantized
fields (John Wiley and Sons, New York, 1980).
\bibitem{Lif-Khal}
E.M. Lifshitz and I.M. Khalatnikov, Adv. Phys. 12 (1963) 185.
\bibitem{Berezin}
F.A. Berezin, The method of second quantization (Academic, New York, 1966).
\bibitem{Faddeev}
L.D. Faddeev and A.A. Slavnov, Gauge fields: introduction to quantum theory
(Benjamin, New York, 1980).
\bibitem{Hall-Hawk}
J.J. Halliwell and S.W. Hawking, Phys. Rev. D 31 (1985) 1777.
\bibitem{Laflamme}R. Laflamme, Phys. Lett. B 198 (1987) 156.
\bibitem{Barv}A.O. Barvinsky, Phys. Rep. 230 (1993) 237.
\bibitem{we-annals}A.O. Barvinsky, A.Yu. Kamenshchik and I.P. Karmazin,
        Ann. Phys. (N.Y.) 219 (1992) 201.
\bibitem{reduc}A.O. Barvinsky, Phys. Rev. D 50 (1994) 5115.
\bibitem{reduc2}A.O. Barvinsky, Phys. Lett. B 428 (1998) 322;
Nucl. Phys. B 520 (1998) 533.
\bibitem{Kam-Mish}A.Yu. Kamenshchik and I.V. Mishakov, Int. J. Mod. Phys. A 7
        (1992) 3713; Phys. Rev. D 47 (1993) 1380; 49 (1994) 816.
\bibitem{Thorne}R.C. Thorne, Phil. Trans. Roy. Soc. London 249 (1957) 597.
\bibitem{Grad-Ryz}
I.S. Gradstein and I.M. Ryzhik, Table of integrals, series, and products
(Academic Press, New York, 1980).
\bibitem{Vach-Vil}T. Vachaspati and A. Vilenkin, Phys. Rev. D 37 (1988) 898.
\bibitem{we-tun}A.O. Barvinsky and A.Yu. Kamenshchik,
Int. J. Mod. Phys. D 5 (1996) 825.

\end{thebibliography}
\end{document}